 \documentclass[final,3p,times,twocolumn]{elsarticle}

 \usepackage{graphicx}
 \usepackage{epsfig}

\usepackage{amssymb}

 \usepackage{lineno}

\journal{Nuclear Instruments and Methods in Physics Research Section A}

\begin{document}

\begin{frontmatter}



\title{Properties and Performance of Two Wide Field of View Cherenkov/Fluorescence Telescope Array Prototypes}


\author[ihep]{S. S. Zhang\corref{cor}}
 \ead{zhangss@ihep.ac.cn}
 \cortext[cor]{Corresponding author. Institute of High Energy Physics, CAS, Beijing 100049. Tel.: +86-010-88236035}
\author[ihep]{Y. X. Bai}
\author[ihep]{Z. Cao}
\author[ihep]{S. Z. Chen}
\author[ihep]{M. J. Chen}
\author[ihep]{Y. Chen}
\author[hebei]{L. H. Chen}
\author[ihep]{K. Q. Ding}
\author[ihep]{H. H. He}
\author[ihep]{J. L. Liu}
\author[ihep]{X. X. Li}
\author[ihep]{J. Liu}
\author[ihep]{L. L. Ma}
\author[ihep]{X. H. Ma}
\author[ihep]{X. D. Sheng}
\author[ihep]{B. Zhou}
\author[ihep]{Y. Zhang}
\author[ihep]{J. Zhao}
\author[ihep]{M. Zha}
\author[ihep]{G. Xiao}

\address[ihep]{Institute of High Energy Physics, CAS, Beijing 100049.}
\address[hebei]{Heibei Normal University, China, Heibei 050016.}


\begin{abstract}

    A wide field of view Cherenkov/fluorescence telescope array is one of the
main components of the Large High Altitude Air Shower Observatory
project. To serve as Cherenkov and fluorescence detectors, a flexible
and mobile design is adopted for easy reconfiguring of the telescope array.
Two prototype telescopes have been constructed and successfully run at
the site of the ARGO-YBJ experiment in Tibet.
The features and performance of the telescopes are presented.

\end{abstract}


\end{frontmatter}

\linenumbers

\section{Introduction}

The energy spectrum of primary cosmic rays spans almost 12 orders of
magnitude, from $10^{9}$~eV to $10^{21}$~eV, and can be well fitted
by a simple power law except in several small energy regions. A region
 called the ``knee"  of the spectrum existing at around $10^{15}$~eV
is one of these regions where the spectrum becomes
steeper at higher energy side. Many experiments have observed this
phenomenon; however, controversial arguments on
its origin persist because of limited discrimination power on the primary cosmic ray
composition and ambiguities in nucleus-nucleus interaction modeling.
These two aspects are closely related to each other. Modern balloon
borne experiments, such as ATIC~\cite{ATIC} and CREAM~\cite{CREAM}, have efficiently
measured the energy spectra of individual elements at the top of the atmosphere. The
energy spectra for all nuclei are measured up to $\sim$100~TeV which
is not far from the ``knee". Because the detector area is
constrained by the payload, the spectrum measurement has to be extended
to a higher energy using a ground based air shower
detector array. The spectrum should initially be measured well
below 100~TeV to create an overlap with the balloon experiments which
serve as absolute calibrations for the ground-based techniques.
Identifying the individual components of cosmic rays continues to be a major challenge
in ground-based experiments.  Multiple parameter measurements on an air shower seem to be a plausible approach.  The ultimate goal is to separate individual species out of the total observed-event samples and measure a clear individual ``knee" for every single species, enabling the discovery of the origin of the ``knee". As one of the major scientific goals of the Large High Altitude Air Shower Observatory (LHAASO) project~\cite{LHAASO-caozh,LHAASO-hhh}, the energy spectrum for a separated composition will be measured at energies above dozens of ~TeV. To tag each primary particle that causes an air shower, the atmospheric depth of the shower maximum should be measured as one of the important parameters.  The wide field of view Cherenkov/fluorescence telescope array
(WFCTA), one of the components of the LHAASO project, is designed to accomplish this goal.

A portable design of WFCTA telescopes is adopted to maximize the flexibility of changing the configuration of the array of telescopes.  The elevations, pointing directions, and locations of the telescopes are then easily reconfigured. This is one way of using the same telescopes to serve as both fluorescence and Cherenkov detectors.  In the fluorescence detector, the telescopes are tilted down to a horizontal position. In such an operational mode, which is analogous to the HiRes experiment~\cite{Hires-detector}, most of the Cherenkov photons are avoided except those that are scattered onto the field of views (FOVs) of the telescope, such as in the fluorescence detector of the telescope array experiment~\cite{TA-exp} and the fluorescence detector of the Pierre Auger experiment~\cite{auger-fluorescence, auger-spectrum}.  Only the fluorescence light from the shower is collected together with the scattered Cherenkov light to trigger the telescopes. This requires showers having much higher energy, usually above 100 PeV, such as in the HiRes prototype experiment~\cite{Hires-hybrid}, because the fluorescence light by a single electron is considerably weaker and  isotropic. In the Cherenkov detector, the telescopes run in high elevation mode to directly measure Cherenkov light from the showers, similar to what was done in the Dice experiment~\cite{dice-exp}. A Cherenkov light radiation provides considerably more photons along the shower axis that are useful for lowering the shower energy.

In 2007, two prototype Cherenkov telescopes~\cite{CRTNT-caozh,CRTNT-hhh} were deployed at Yangbaijing (YBJ) Cosmic Ray Observatory near the ARGO-YBJ experiment~\cite{ARGO-detector}. Moreover, two WFCTA telescopes have been successfully running in Cherenkov mode beginning August 2008.  To date, millions of cosmic ray events that simultaneously trigger the telescopes and the ARGO-YBJ detector carpet array have been collected. An analysis of these events is carried out to study the performance of the telescopes.  Detailed descriptions of the telescopes and the analysis of the findings are presented in this paper.

Several details about the apparatus are presented in Section 2. The detector calibration is then discussed in Section 3. The test-run of the two telescopes and results are reported in Section 4 including summaries on the detector performance. The conclusions drawn are provided in the last section.
\section{Apparatus}
The  two  prototype  telescopes  are  deployed  near the ARGO-YBJ carpet detector array at a longitude of
$90.53^{\circ}$E, and a latitude of $30.11^{\circ}$N and 4300 m a.s.l. One
telescope is about 25~m away from the west side of the ARGO-YBJ array. The other is also 25~m away from the
south side of the array with  separation distance between the two telescopes is 50~m. Each telescope has an
FOV of $14^{\circ}$ in elevation by 16$^{\circ}$ in azimuth. The focal
plane camera is made of a 16$\times$16 photomultiplier tube (PMT) array, and the pixel size is approximately $1^{\circ}$. Because both telescopes are tilted up to 60$^{\circ}$ pointing in the
same direction, they can be operated in stereoscopic mode, i.e.,  showers striking an area covered by the telescopes will be seen simultaneously.  Since the Cherenkov light from a shower is very concentrated in a forward
region; thus, the telescopes can be triggered by showers coming within a
cone of approximately 8$^\circ$ with respect to the main optic axes of
the telescopes.

\begin{figure}[!t]
\centering
\includegraphics[width=0.5\textwidth]{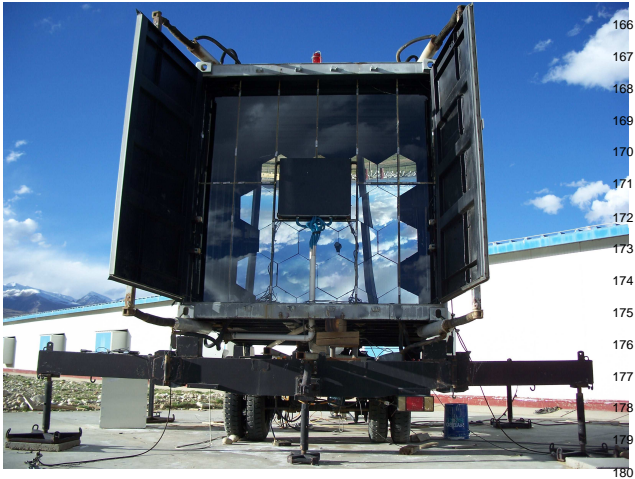}
\caption{Photograph of the telescope with the doors open.} \label{telescope}
\end{figure}

\begin{figure}[!t]
\centering
\includegraphics[width=0.5\textwidth]{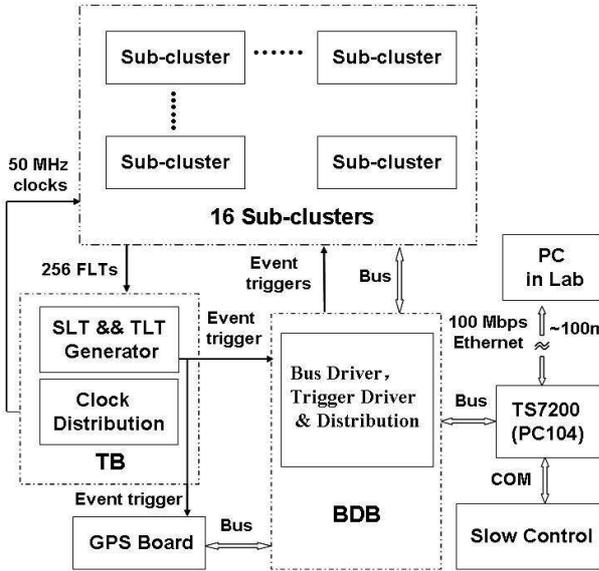}
\caption{Communications diagram of one telescope; for details of the sub-cluster see Fig.\ref{subcluster-double-fig}.} \label{com_diag}
\end{figure}

The entire telescope system is composed of an optic ultraviolet light collector, a focal plane camera, front end electronics (FEE) based on 50-MHz flash analog-to-digital-converters (FADC), data acquisition (DAQ) based  on an  embedded ARM processor and PC104 bus, power supplies for low and high voltages, and a slow control system. Everything is installed in a shipping container with dimensions of 2.5~m$\times$2.3~m$\times$3~m (Fig.\ref{telescope}). Mirrors are mounted at one end of the container
and the camera is located at the other end where the focal planes of the mirrors are. The FEE and DAQ are placed at the back plane of the PMT camera.  A glass window is installed at the entrance aperture to keep dust from entering the apparatus. The container is mounted on a dump-truck frame with a hydraulic lift that allows the container to be
lifted up from 0$^{\circ}$ to 60$^{\circ}$. The mobility of the entire telescope allows for freely switch between configurations of the telescope array for different observational
modes.  The architecture of the electronic data acquisition and the slow control system are shown in Fig.\ref{com_diag}, whereas that of a sub-cluster is shown in Fig.\ref{subcluster-double-fig}. The PMT signals are processed using an analog processing board (AB) and then a digitization board (DB). The first level trigger (FLT) is generated in the DB on a sub-cluster. After the FLT is determined, 256 FLTs are then sent to the trigger board (TB). The second level trigger (SLT) and the third level trigger (TLT) are then determined in the TB. After this, the event trigger from the TB is fanned out by a bus driver board (BDB) and sent back to each of the DB and GPS boards. The data are initially stored in the buffer of the DB when the DB has received
the event trigger, after which the data that include the GPS time
are read using TS7200. Finally, the data are stored in a PC in the laboratory via Ethernet. A detailed description of the detector, divided into the following
8 subsystems, is provided: 1) optics, 2) camera, 3) FEE, 4) trigger system, 5) DAQ, 6) power supply system, 7) slow control system that includes monitoring of everything, and 8) calibration.

\subsection{Optics}
A 4.7~$m^2$ spherical aluminized mirror, composed of 20 hexagon-shaped segments, is used as an ultraviolet light collector. Each segment is subjected to
a strict control of the surface quality and their geometrical and optical properties.  The reflectivity is greater
than $82\%$ for light having a wavelength $\ge$300~nm.  The radius
of the curvature of the segments is 4740~mm with a tolerance of $\pm$20~mm.

The size of a light spot on the focal plane where the PMT camera is located is designed to be similar to the size of a pixel.  The sensitivity of a PMT is non-uniform across the surface of the photocathode~\cite{G-xiao}, and small gaps exist between PMTs in the camera. For a spot that is exceedingly larger than the dimension of the gaps and the typical spatial scale of the non-uniformity across the photocathode, light from a specific direction will be shared by a few adjacent PMTs. Thus, effects stemming from the overall non-uniformity of response across the entire camera are reduced. The direction of incident photons can be efficiently determined by simply taking the average of the directions of the registered pixels as weighted by the measured charge in a pixel. Contrastingly, the size of a spot is optimized to be similar to pixel size to avoid sharing of incident photons by an excessive number of pixels.

Because of the abbreviation of the spherical reflector, the spot size changes across the focal plane in a rather
large FOV. Large comas also occur at large off-axis angles.  The uniformity of the spot size over the
camera is also optimized by locating the camera slightly away from the focus. A distance of 2305~mm between the mirror and the camera is eventually set according to a detailed ray-tracing calculation.

Each mirror segment is mounted on a spherical steel frame with three adjustable screws. The pointing orientations of all segments are adjusted toward the geometric center of the curvature. Using a laser beam, the orientations are calibrated to be less than 7.6 arc seconds from the nominal direction.
\subsection{PMT Camera}

Photons in a spot at the focal plane are recorded by a camera composed of 256 pixels which are 40~mm Photonis
hexagonal PMTs (XP3062/FL). 
The camera is enclosed in a box and mounted on a frame at a distance of 2305~mm from the spherical mirror.

The PMTs in a telescope are operated at a gain of $6\times10^5$. To achieve the best uniformity at hardware level, a resistor is placed between the high voltage power supply and each PMT base to compensate for the gain difference between PMTs. The gains (G) for all PMTs and their responses to the supplied voltage, i.e., $G\propto V^{-\beta}$, are calibrated. All $\beta$ values are measured and recorded in a database in the laboratory for future use. The PMTs are then sorted according to their gains and grouped into two classes. For instance, the working voltages of the PMTs of one of the telescopes distribute 1088 V to the telescope in a common power supply with a variance of 73.97 V. The average $\beta$ of these tubes is 5.9 with a variance of 0.6.

The 16 PMTs, as an integrated unit, are soldered on a high voltage board (HVB) that distributes a negative
high voltage to all cathodes and dynodes of the PMTs. The voltage division scheme is recommended by Photonis
to yield the maximum gain of the tubes. The maximum output current of the PMTs allows a range of 3.5 orders
of magnitude in which the non-linearity of all tubes is
less than 8\% according to a calibration in the laboratory~\cite{G-xiao}. The anode signals are finally DC coupled to the FEE.

\begin{figure*}[!t]
\centering
\includegraphics[width=0.4\textwidth]{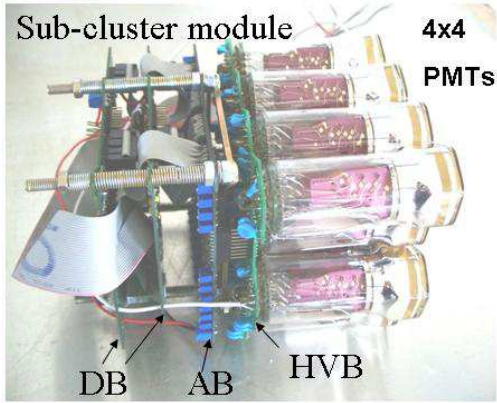}
\label{subcluster}\hfil
\includegraphics[width=0.5\textwidth]{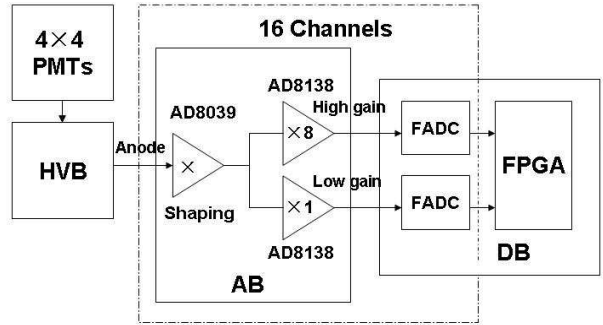}
 \label{subcluster_shematic}
\caption{Photograph of a sub-cluster (left) and schematic of the sub-cluster (right).}
   \label{subcluster-double-fig}
 \end{figure*}

\subsection{FEE and Digitization}

The FEE is located behind the HVB, which is composed of an AB and two DBs, to avoid a long distance transmission of the analog signals. Such a module is referred to as a sub-cluster (see the left figure of Fig.\ref{subcluster-double-fig}) in this paper. A block diagram on the procession of signals from the PMTs to the on-board data storage chip is shown in the right figure of Fig.\ref{subcluster-double-fig}. The signal of each PMT is transmitted to the first amplifier on the AB for noise filtering and pulse stretching. Upon division into high or low gain channels, the signal is then sent to the DB for digitization and further processing. The rest of this section describes the processing procedure in detail.

\subsubsection{Analog Processing Board}

Each AB has 16 channels for the signals from 16 PMTs on HVB alone. Each channel has a shaping circuit and amplifiers for low or high gains. The board has four main functions, namely:

\begin{itemize}
\item low-pass filter of 20 MHz,
\item expanding narrow pulses,
\item dual-gain system for covering a dynamic range over 3.5 orders of magnitude,

\item and receiving PMT signals from HVB and performing a single-ended-to-differential conversion.
\end{itemize}

\begin{figure}[!t]
\centering
  \includegraphics[width=0.5\textwidth]{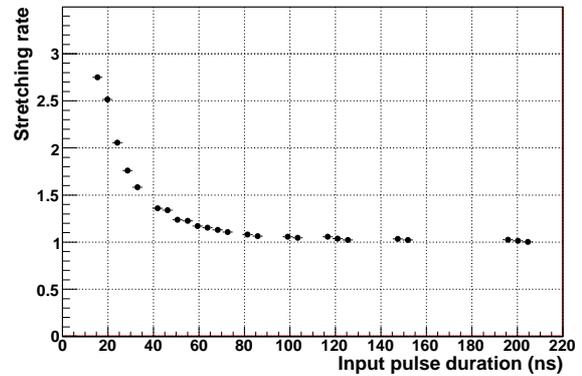}
  \caption{Stretching ratio as a function of the input pulse duration.}
\label{strech-rate}
\end{figure}

The anode signals are fed to a four-pole low-pass filter based on an AD8039 through DC coupling. Cherenkov photons generated by all shower particles move at almost the same speed as the charged particles; thus, all photons generated over the entire shower development beginning from the top of the atmosphere hit the cathodes at almost the same time. The duration of the pulse lasts only a few nanoseconds, which is shorter than the typical response time of the PMT, e.g., a pulse duration generated by a single-photo-electron (SPE) is approximately 12~ns on average. Therefore, such a narrow pulse has to be stretched (e.g., at $>50$~ns) to be able to measure the charge at a sampling rate of 50 MHz, as preselected for the FADCs on DBs. Taking into account the optimized bandwidth of the noise filter, the stretching ratio is selected in such a manner that the narrower pulses are stretched further, and pulses wider than 120~ns are essentially not stretched. In Fig.\ref{strech-rate}, the stretching ratio is plotted as a function of the pulse duration. According to this, the original waveform is reconstructible, and the timing of the pulse can be corrected with an uncertainty less than the duration of the stretched pulses.

The amplification of the PMT signals with a gain of 2.67 is another feature of the shaping circuit. The value of the gain is also optimized together with the filter and the stretcher. The AD8039 has a sufficiently high speed
(350 MHz), low power dissipation, low cost, low noise, low distortion, and a nonlinearity less than 1\% in a range of the input from 0.5~mV to 800~mV, which nearly fits the entire range of the PMT signals.

To maintain good linearity over a wide dynamic range of 3.5 orders of magnitude in charge, a dual gain system is designed. Signals coming out of the AD8039 are split into two channels and are then separately amplified by two AD8138. With such a design, considerable flexibility in obtaining different resolutions of the charge measurements for pulses with different pulse heights is achievable by choosing the range covered by the high gain channel. In this paper, a ratio of the gains of 1:8 is selected as a result of the optimization between the dynamic range and the resolution in the charge measurement. The nonlinearity of AD8138 is less than 2\% at both gains of 1 and 8 in the entire range of the input signals.

\begin{figure}[!t]
\centering
  \includegraphics[width=0.5\textwidth]{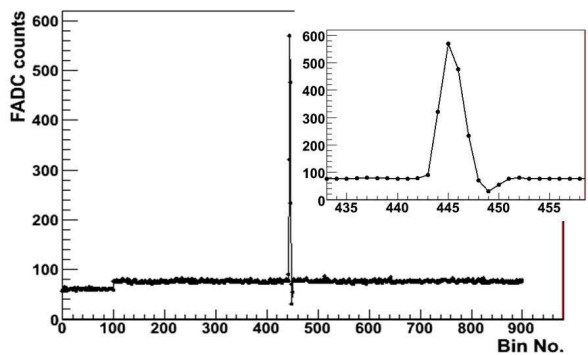}
  \caption{A Cherenkov signal with sky background; the first 100 bins are low gain background and the remaining 800 bins are high gain background and the Cherenkov signal. The Cherenkov signal is in the current frame (from the 300th bin to the 600th bin).} \label{Cherenkov-signal}
\end{figure}

Another excellent feature of the AD8138, the conversion between a single-ended input and differential output for instance, makes it the best choice for an amplifier. Superposing an offset to the output of the AD8138
as a pedestal of the signal before feeding into the positively-polarized FADC is also a convenient approach.
To manage possible undershoot of the pulses (see Fig.\ref{Cherenkov-signal}), a pedestal is preset higher than the undershoot.

\subsubsection{Digitization Board}

Each DB has 8 pixel channels, which includes 16
FADC modules managed by two field programmable gate array (FPGA) modules. The 10-bit FADCs, AD9215,
has a good cost-performance ratio with a linearity better than 2\% over a range of 40 to
900 counts. However, it is not sufficient to cover the required dynamic range of the detector. Two FADCs
are then used for the dual gain system. At an input range of 2 V, the FADC continuously digitizes analog signals
with a sampling rate of 50 MHz (20 ns/bin) and a resolution of 2 mV per count. In the dual gain system, a
range of 0 to 500 photoelectrons with a resolution of 1.7 counts per photoelectron is set for the high gain channel,  whereas a range of 0 to 4000 photoelectrons with a resolution of 0.21 counts per photoelectron is set for the low gain channel.

A digitized waveform is collected by an FPGA, Xilinx XC3S1000, and fed into a pipeline with a length of 1500 clock cycles. Every 300 cycles is defined as a frame in which a single channel trigger is formed. Such a long pipeline enables an enduring waveform waiting for global trigger formation and transmission. This pipeline also allows three frames, i.e., previous, current, and post ones, to be recorded once a global trigger is received.  This guarantees storage of a complete wave form regardless of when the pulse starts in the "current" frame.

The FPGA also makes a choice between the signals from high gain and low gain channels when the bit stream flows in. Any signal higher than 900 counts triggers a switch from the high gain channel to the low gain
channel. This way, an effective overall dynamic range of 12 to 13 bits is achieved.

\subsection{Trigger system}

\begin{figure}[!t]
\centering
  \includegraphics[width=0.5\textwidth]{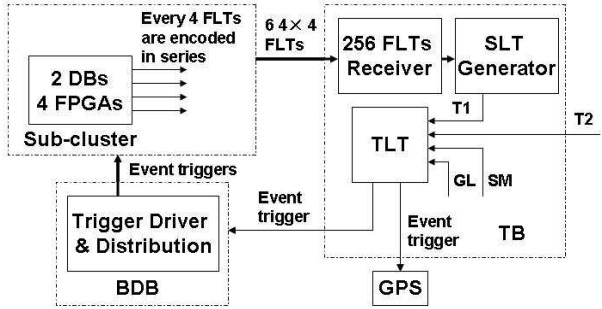}

  \caption{Three-level trigger system.  SM is stereo or mono enable signal, GL is global or local enable signal, T1 is the telescope trigger from telescope No. 1 and T2 is from telescope No. 2.  See detailed explanation of the diagram in the text.}
\label{three-level-trigger}
\end{figure}

A final trigger among the telescopes is determined based on the three trigger levels, namely, the single channel trigger (the lowest level), telescope trigger (the second level), and event trigger (the highest level). The topology of the three-level trigger algorithm is outlined in Fig.$\ref{three-level-trigger}$, with corresponding details discussed in following subsections.

\subsubsection{First Level Trigger in a Single Channel}

The FLT is formed in the FPGA located on the DBs provided that the signal-to-noise ratio in a window is greater than a given threshold (e.g., 4 as a typical value). The width of the window is predetermined for corresponding observation modes (e.g.,  8 bins for the Cherenkov light signals). Running over the entire frame of 300 cycles once in a bin, 293 sums of FADC counts in the windows (denoted as WINSUM) are produced. The average of WINSUM is calculated using these WINSUMs avoiding the maximal WINSUM and 18 WINSUMs on each side of the maximum. The standard deviation ($\sigma$) of the WINSUM around the average serves as a measure of the background fluctuation. If the maximum of WINSUMs exceeds the average by $n\sigma$,  the FLT is formed, where $n$ is preselected from a list of $\sqrt{12}$,
$\sqrt{16}$, $\sqrt{24}$, $\sqrt{32}$ and $\sqrt{40}$ before starting the run.
 All trigger signals from 64 FPGAs are  encoded into bitmaps and transmitted to the
Trigger Board (TB) in parallel (see Fig.$\ref{three-level-trigger}$).

The frame size of 300 bins is selected for a full scale array with more than 24 telescopes covering an area of 1~km$^2$, over which a highly inclined air shower takes several microseconds to cross in the fluorescence light observational mode. Such a frame is large enough to fully contain the entire shower; it is also large enough for Cherenkov light. All Cherenkov photons arrive at the telescopes at almost the same time (few ns). Therefore, an air shower signal appears only within 5 or 6 bins after shaping and the night sky background is measured in the rest of the frame (see Fig.$\ref{Cherenkov-signal}$). Such continuous measuring of the sky background is highly useful not only in estimating the signal-to-noise ratio but also in monitoring the transparency of the atmosphere using well-known bright stars in the field of view.

\subsubsection{Second level trigger for a single telescope}

\begin{figure}[!t]
\centering
  \includegraphics[width=0.5\textwidth]{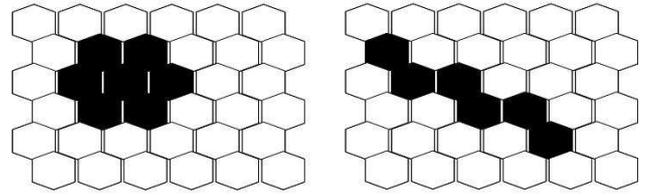}
  \caption{Two typical patterns in the second level trigger.}
\label{patterns}
\end{figure}

A trigger for a telescope registered by an air shower (denoted as second level trigger or SLT) is formed when a specific pattern of triggered PMTs corresponding to a possible air shower is found in the camera of a telescope.  A pattern recognition technique, developed in the Pierre Auger experiment~\cite{auger_patern} that is operated in a FPGA located on the TB, is applied.  Numerous patterns, such as a fully filled circle formed with one hexagonal pixel surrounded by six others, or a straight line formed by six aligned pixels, are pre-loaded into the FPGA as a look-up table. Once all 256 FLT signals are collected, the FPGA matches all the pre-stored
patterns with the observed one within a $6\times6$ matrix
of pixels. Moreover, it keeps such a box running throughout the entire PMT camera with a step of one row or one column. The SLT is formed as long as any one of the pre-stored patterns is matched. According to the simulation for air showers, mainly two types of patterns exist. Fluorescence light images of air showers seen from several kilometers away tend to form line-shaped patterns on the camera, whereas Cherenkov light images of showers hitting the telescope head-on tend to form round-shaped patterns, as shown in Fig.\ref{patterns}. There are 16 round-shaped patterns and 729 line-shaped patterns in a $6\times6$ box.

To speed up the formation of SLT, the pattern comparisons in the box are conducted in parallel. All 121 bitmaps of the boxes are generated by sliding the box are done in parallel. All 121 bitmaps of the boxes
are generated by sliding the box throughout the camera. A pattern comparison algorithm is then carried out among the 121 bitmaps.  The telescope trigger (SLT) is formed in 123 clock cycles, i.e., 2.46~$\mu$s.

\subsubsection{Third level trigger for entire array of telescopes}

An event trigger for the entire array of telescopes (denoted as the third level or TLT) is generated using one
of the FPGAs used for SLT. For the two
prototype telescopes, only two modes, namely, stereoscopic and monocular observation of showers, exists if they are configured in such a way that the two telescopes have a maximal overlap of the FOV. In stereoscopic mode, the two telescopes are required to simultaneously observe a shower. In monocular mode, each of the two telescopes can trigger the entire site. The operational mode should be selected at the beginning of a run by assigning two controlling parameters, SM and GL, as marked in Fig\ref{three-level-trigger}.

The two telescopes communicate with each other through a two 80 m coaxial cables by sending telescope triggers (SLT) out and receiving the event trigger (TLT). Such a trigger function is also useful for receiving an external trigger and delivering a trigger to other co-sited detectors. This makes the telescope an open and modularized system.

\subsection{DAQ system}

The philosophy of the design of the Cherenkov telescope DAQ is integrated and compact so that the entire DAQ can be arranged on the backboard of the focal plane camera. It allows for the maximal mobility of the telescope for flexibility in switching between configurations. Because of the limited space, a low power consuming and compact-embedded online computer based on an ARM processor, industrial standard PC104 bus, and flash disk are selected as the backbone of the DAQ system, eliminating the need for moving parts such as the CPU fan and hard drives. All of the components are integrated on a bus driver board, which bridges the data storage disk and the 32 DBs that are connected by flat cables. Because the event rate is not extremely high, i.e., around 1 Hz, a band width of 14 MHz for a PC104 bus is sufficient for transferring every bit of the 6~$\mu$s long wave forms to the disk for all 256 channels. It is also extremely advantageous for system debugging. Each telescope is an independent detector with a complete DAQ system on board. The communication with the rest of the experiment is through a 10/100 Mbps Ethernet.

Each telescope has its own independent DAQ, developed using C++ language under Linux. Operating in a polling mode instead of an interrupt mode, the DAQ checks the interface status regularly for whether the data are ready or not. If so, the data in the buffer are read and immediately stored in the disk. Before a night shift ends, the data in the hard disk are moved to the Institute of High Energy Physics (IHEP) in Beijing for further analysis.

\subsection{Power supply, slow control, and status monitoring}
The two power supplies with +7 V (maximum current 80 A) and -7 V (maximum current 20 A) are installed in
the container. Regulators are used at each AB and DB to further stabilize the voltage. An adjustable HV power supply with a maximal output of -2000 V and 100 mA is used for each telescope.

One of the difficulties at high altitudes is the heat dissipation of the power supplies and electronics enclosed in a metal box. The total power consumption is about 50 A at +7 V and 16 A at -7 V. A forced cooling system is necessary in the prototype experiment to improve the heat dissipation of the entire camera and the power supplies.

The entire telescope system is powered by an uninterrupted power supply with a sufficient battery
backup. It protects the telescopes from damage when a blackout occurs.

The detector is designed to work in remote control mode, including the opening and closing of the doors, turning
the power supplies on/off, enabling/disabling the high
voltage (HV) power supply and low voltage (LV) power supplies, and switching the UV LEDs on/off for detector calibration. All kinds of controls are realized using the on-board computer through a COM port. Carefully monitoring the status of each telescope is necessary, including the door status, voltages, and temperatures at different places (e.g., enclosed areas inside the camera, backboard of the PMT camera, and inside the
housing of the UV LED). All parameters are measured and recorded through an 8-channel 12-bit analog-to-digital-converter on the board of the embedded computer, TS7200.

All of the controls are performed through a user interface running at the embedded computer connected through Ethernet from an operational center, 3000 km away in IHEP in Beijing.  A small portion of the data can be copied to display the event during the operation.

\section{Calibration}
\subsection{Method of calibration}

An accurate shower reconstruction requires a converting factor from a pulse area in terms of FADC counts to the number of photons for each pixel. Therefore, the absolute calibration of the detector response is essential. To achieve this objective, having an accurate knowledge about PMT cathode effective areas, cathode quantum efficiency, PMT gains, amplifier gains, and digital converting factors, is necessary. Measuring all these effects
item by item is difficult. In this paper, a method similar to the HiRes experiment~\cite{Hires_calibration}
and Pierre Auger experiment~\cite{auger_calibration} calibration procedure, which considers the entire effect, is applied to measure the overall response of each pixel. The procedure is discussed in the paragraphs that follow.

Being mounted at the center of the mirror, a UV-LED (375 nm) light source with a diffuser is used for the calibration of PMTs in the camera by beaming nearly uniform light to every pixel. The LED light density on the camera surface is calibrated using a pre-calibrated probe detector located beside the camera. The calibration of all pixels in the cameras is performed twice a day, i.e., before and after the daily operation. The crucial part is the measurement of the absolute number of UV photons at each PMT cathode from a pulse emitted by the LED, which is performed in two steps.

First, we move one PMT over all the places on the frame of the PMT camera to measure the uniformity of the LED light density on the camera surface. This has to be accomplished before the PMT camera is installed. The light density is a function of the polar angle, $\theta$, in a form of $cos^4\theta$, where $\theta$ is the angle between the connection from the LED to the PMT location and the perpendicular connection from the LED to the PMT camera surface.  The variation from the center to the corner of the PMT camera is within 7$\%$.

In the second step, the light density is measured using a pre-calibrated probe consisting of two XP3062 PMTs with the same FEE and DAQ as the two telescopes. The only difference is that the two PMTs in the probe are operated at a very high gain so that the single photoelectron can be measured. Therefore, the gain of the probe, $G^{probe}$, can be calibrated at any time.  Then, the absolute calibration of the probe is performed at the HiRes lab at the University of Utah, USA, by comparing the response of a hybrid photo diode (HPD) pre-calibrated at NIST~\cite{HPD-calibration} to the same light source.  The probe and HPD are located side by side in front of  a UV-LED at 355 nm. Using the HPD, we measure light density $I_{U}$ (number of photons per square millimeter) from the LED. We also measure pulse area $F^{probe}_{U}$ using the probe simultaneously. At the operational site in Tibet, the light density from the LED mounted on the center of the mirror is calibrated to be
$I_{T}=I_{U}\frac{F^{probe}_{T}}{F^{probe}_{U}}\frac{G^{probe}_{U}}{G^{probe}_{T}}$,
where $U$ and $T$ stand for Utah and Tibet, respectively.

Applying the knowledge obtained from the two steps, we have determined both the light density in front of the cathode and the pulse area in terms of the FADC counts for each PMT in the camera.  Finally, the overall converting factor for a pixel between the number of photons reached to the surface of the camera and corresponding pulse area measured by the FADC counts behind the pixel is
\begin{equation}\label{calib}
C_{375}^{camera}=\frac {F_T^{camera}}{I_{T}A_{PMT}}\frac{cos^4\theta^{probe}}{cos^4\theta^{camera}}  ,
\end{equation}
where $A_{PMT}$ is the geometric area of the PMT cathode, and $C_{375}^{camera}$ is the calibration constant for a pixel. The subscript $375$ indicates that the calibration is done using UV light at 375~nm. The unit of the calibration constant is FADC counts per photon.
$\theta^{probe}$ and $\theta^{camera}$ represent the angular locations of the probe and the pixel in the
camera, respectively. A further correction according to the wavelength dependence of the quantum efficiency
of PMTs is applied in the operation for the cosmic ray observation.


\subsection{Result of the calibration}

The probe was calibrated at the HiRes lab. The calibration results are shown in table.\ref{table-probe}; the gain of the probe is 85.7$\pm 0.3$ FADC count per SPE. The second column
shows the lower LED photon density and the third column
shows the higher LED photon density that coincide with one another.

 \begin{table}[!h]
  \caption{The calibration results of the probe.}
  \label{table-probe}
  \centering
  \begin{tabular}{|c|c|c|}
  \hline
    $I_{U}$(photons/$mm^2$) & 0.359 $\pm 4.8\%$ & 0.538 $\pm 4.8\%$ \\
    \hline
    $F^{probe}_U$(FADC count)& 8959 $\pm 4$ & 13650 $\pm 5$ \\
  \hline
  \end{tabular}
  \end{table}

\begin{figure}[!t]
\centering
  \includegraphics[width=0.5\textwidth]{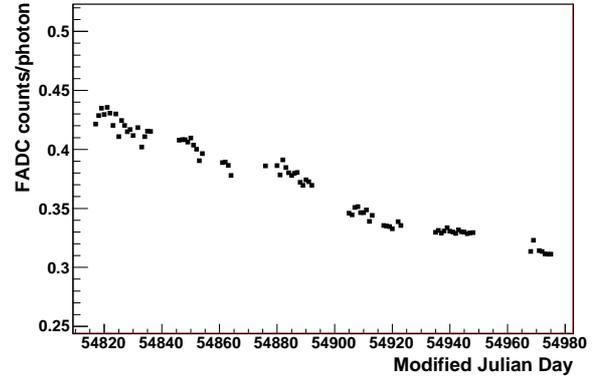}
  \caption{Absolute gain of one telescope from Dec. 2008 to May 2009, as a function of time.} \label{gain-day}
\end{figure}

 The corresponding calibration constant for one of the telescopes is presented in Fig.\ref{gain-day}. An average of the calibration constants of all pixels in the camera is plotted with respect to time from December 2008 to May 2009. As mentioned above, the calibration constant is monitored every observational day. The systematic uncertainty of the calibration constant is estimated to be 7\%. The downtrend of the constant shown in the figure indicates obvious decreases of gains in all newly produced of PMTs. The transmission of the glass window and reflectivity of the mirrors are not take into account in the above calibration. These two effects will be monitored in so called end-to-end calibration using nitrogen laser in future.

\section{Performances and Results from Test Run}

\subsection{Test Run Information}

The two telescopes began recording cosmic ray data in August 2008.  Furthermore, both monocular and stereoscopic modes have been tested. About 500,000 coincidental events with the ARGO-YBJ experiment in stereoscopic mode and 700,000 coincidence events in monocular mode have been collected up to January 2010. The average trigger rate is about 0.5 Hz in stereoscopic mode and 0.7 Hz in monocular mode.

\subsection{Off-line Coincidence with the ARGO-YBJ Experiment}
\label{event-matching-data}

\begin{figure}[!t]
\centering
  \includegraphics[width=0.5\textwidth]{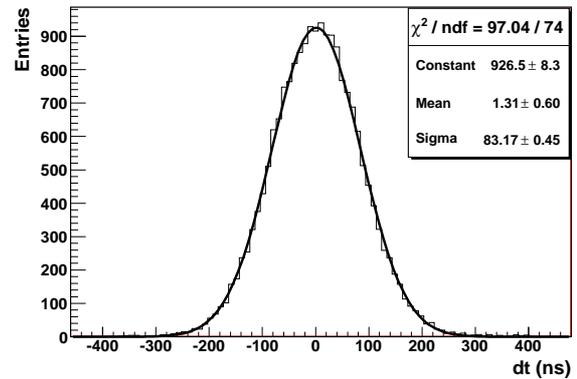}
  \caption{The difference time between WFCTA and ARGO in coincidence event.}
\label{time-dif}
\end{figure}

All WFCTA telescopes and the ARGO-YBJ experiment recorded the cosmic ray arrival time based on a GPS. A time window of 8~$\mu$s containing a Cherenkov event is searched for coincidence with the ARGO-YBJ event stream, which is about 4~kHz. A difference between the recorded event time by the two experiments for a matched event is typically less than 100 ns (Fig.\ref{time-dif}).

\begin{figure}[!t]
\centering
  \includegraphics[width=0.5\textwidth]{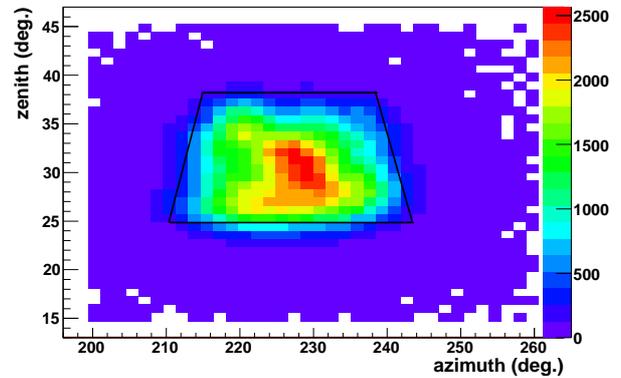}
  \caption{Distribution of coincidence events are seen by ARGO and Cherenkov telescopes simultaneously over
  zenith and azimuth angle. About $85\%$ of coincidence
events are located in the Cherenkov telescope FOV (trapezium). The image of the rest
of the $15\%$ of coincidence events is partially in the Cherenkov telescope FOV.}
\label{the-phi}
\end{figure}

\begin{figure}[!t]
\centering
  \includegraphics[width=0.5\textwidth]{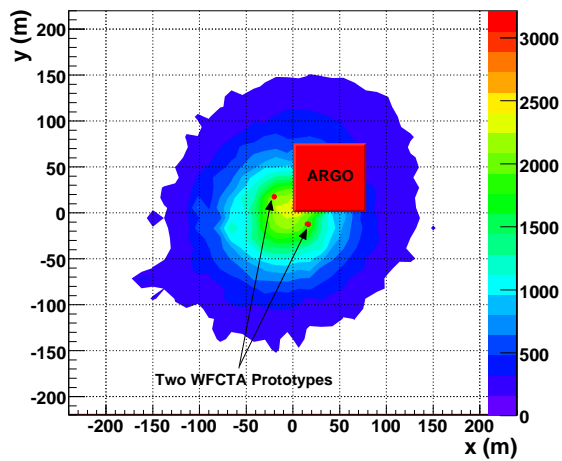}
  \caption{Distribution of shower core. Two Cherenkov telescopes (two dots) and the ARGO-YBJ experiment array (rectangle) are also marked in the figure.}
\label{core}
\end{figure}

\begin{figure}[!t]
\centering
  \includegraphics[width=0.5\textwidth]{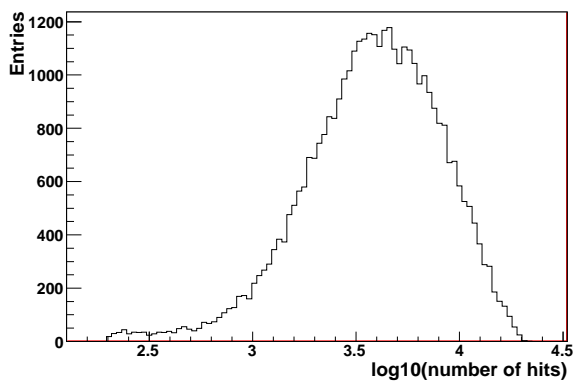}
  \caption{Distribution of ARGO nhit of these coincidence events, whose reconstructed core is located in the ARGO-YBJ cluster array.} \label{nhit-raw}
\end{figure}

For a coincidence event, the shower geometry is measured using the ARGO-YBJ detector. The distribution of shower arrival directions is shown in a 2-dimensional map,  i.e., zenith angles versus azimuth angle, as in Fig.\ref{the-phi}. Approximately $85\%$ of coincident events occur inside the FOV of the Cherenkov telescopes marked by the trapezium. The rest of the 15\% of events have their images partially seen by the telescopes and sufficiently trigger the telescopes.  The shower core distribution is shown in Fig.\ref{core}. The ARGO-YBJ experiment array and the two Cherenkov telescopes are marked as the rectangle and two dots in the figure, respectively. For events that have the reconstructed cores inside the ARGO-YBJ array, the shower parameters, such as the number of hits on the carpet detector ($N_{hit}$) and shower geometry, are well measured.  A distribution of $N_{hit}$ of these events is plotted in Fig.\ref{nhit-raw}. According to the number of hits as a function of the primary energy of proton from
ARGO-YBJ detector~\cite{guoyq}, the mode energy of protons is about 40 TeV. This
estimates the threshold of two Cherenkov telescopes in stereoscopic mode.

\begin{figure}[!t]
\centering
  \includegraphics[width=0.5\textwidth]{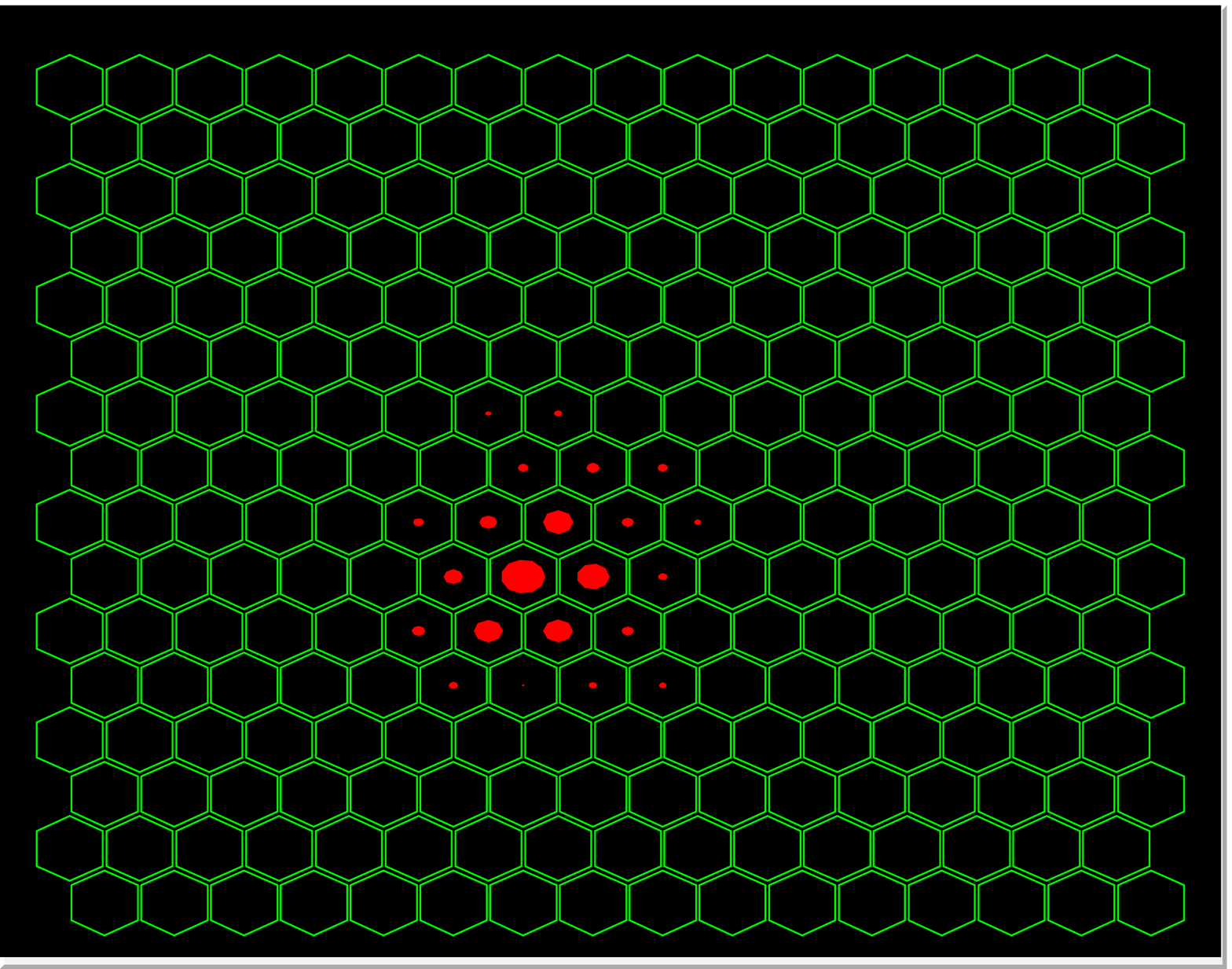}
  \includegraphics[width=0.5\textwidth]{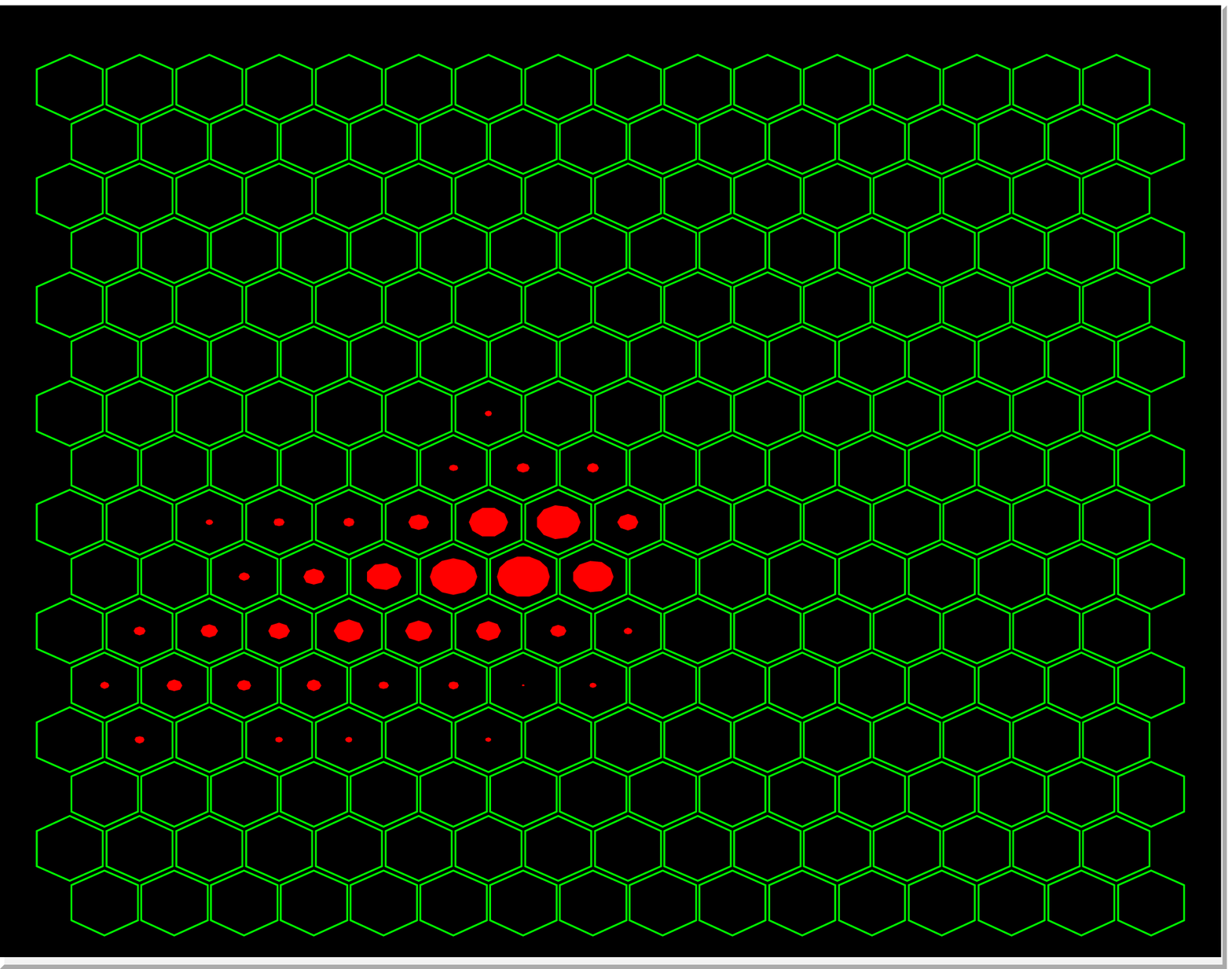}
  \caption{Two Cherenkov events; the Rp of the tope event is 10 m and that of the bottom event is 173 m; the image of the bottom event has a longer tail than the top event.}
\label{Cherenkov-event}
\end{figure}

\begin{figure}[!t]
\centering
  \includegraphics[width=0.5\textwidth]{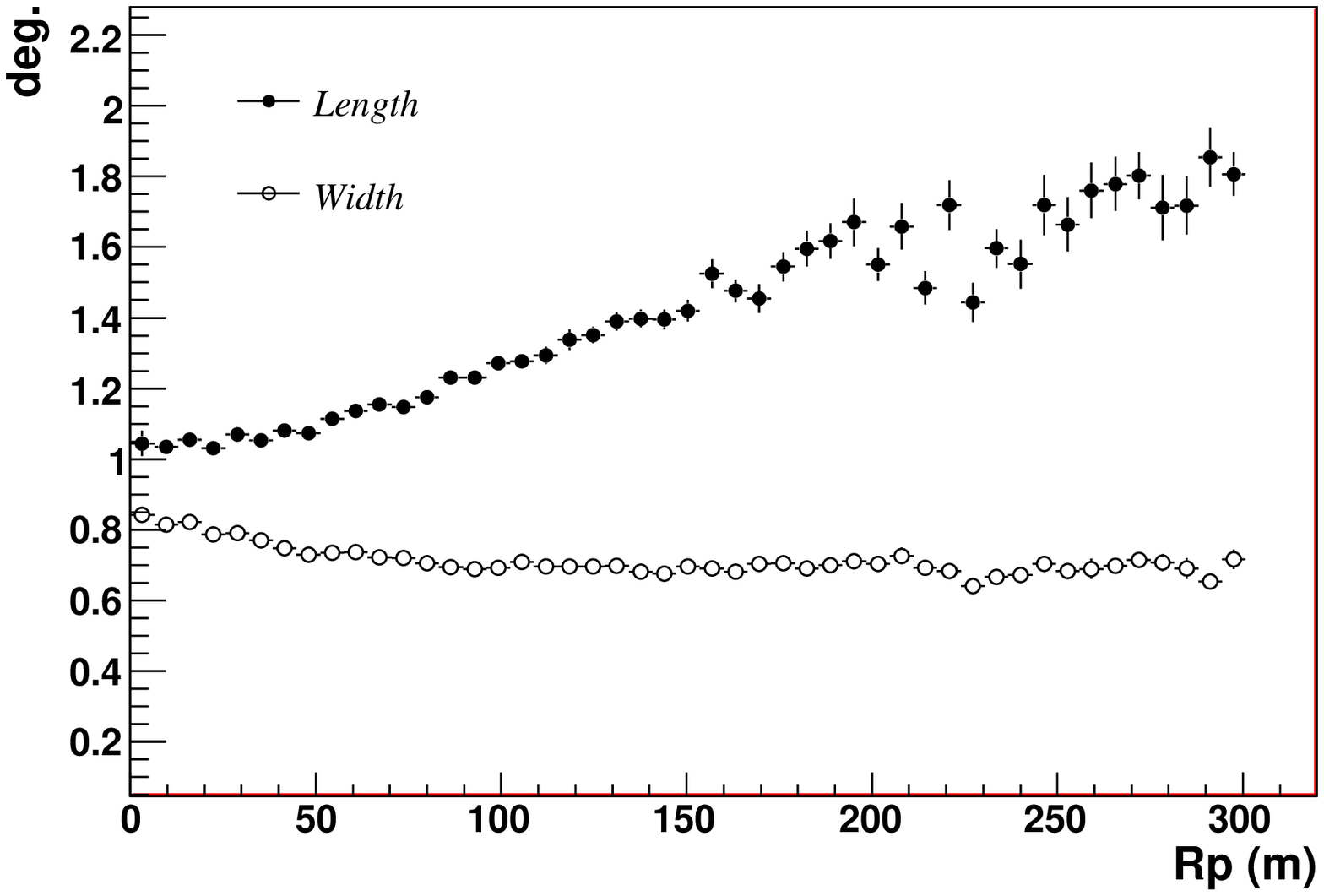}
  \caption{Length (filled circles) and width (open circles) as a function of Rp.}
\label{L-W-Rp}
\end{figure}

In Fig.\ref{Cherenkov-event}, two Cherenkov images are shown in the event display. The top image is caused by an event at a 10 m $R_p$ (the impact parameter of the shower to the telescope) and the bottom image stems from an event at a 173 m $R_p$. The image is clearly more elongated for the farther event. Using the well-defined image parameters created by Hillas~\cite{Hillas}, length and width, this effect is more quantitatively presented in Fig.\ref{L-W-Rp}. The widths of the images seem to be no longer shrinking once showers are sufficiently far from the telescopes (e.g.,  farther than 100 m).  The showers essentially resemble linear patterns.

\subsection{Electronic noisy and sky background}

For a triggering system completely based on the signal-to-noise ratio, such as the WFCTA telescopes, the sensitivity is constrained by the noise level. For a most optimized system, the electronic noise must be negligible compared with the night sky background.  Both of them are measured during the operation as the door is closed and opened, respectively.

\begin{figure}[!t]
\centering
  \includegraphics[width=0.5\textwidth]{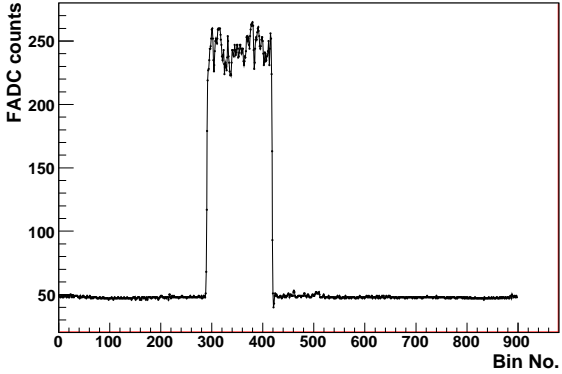}
  \caption{LED calibration for door closed. The low noise level of 1 LSB of the complete electronics is to be
           seen before and after the LED pulse.}
\label{LED-signal}
\end{figure}

The sources of electronic noise include thermal noises in the PMTs, noises from the HV power supply and LV power supplies, noises of amplifiers, and finally from the counting error of FADCs.  The PMT Photonis XP3062/FL has a very low dark current; therefore, the thermal noise level is sufficiently low so that it can be ignored compared with the other sources of electronic noises. The HV power supply has a ripple less than 0.02\% in terms of RMS of rated voltage. This contributes a variation within 0.12\% in terms of the gain of the PMTs. Voltage regulators are used to block possible noises from the LV power supplies.  All amplifiers, AD8138 and AD8039, are selected to generate a very low noise level. The noise from the FADC is about 0.5 FADC counts per tube. Taking into account all additional sources such as the distributing capacity on boards and connectors, the total electronic noise, including PMT and high voltage power supply, is typically about 1.0 FADC counts per tube. It is measured during the calibration with the LED (Fig.\ref{LED-signal}) as long as the signal is avoided.

\begin{figure}[!t]
\centering
  \includegraphics[width=0.5\textwidth]{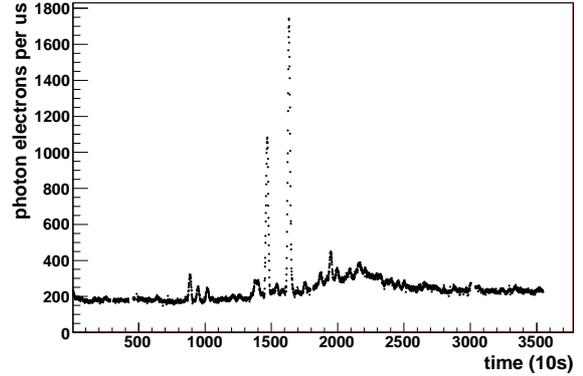}
  \caption{ Sky background is monitored by a PMT in a clear moonless night; each pulse corresponds to a star passing through the FOV of the telescope.} \label{sky-bk}
\end{figure}

\begin{figure}[!t]
\centering
  \includegraphics[width=0.5\textwidth]{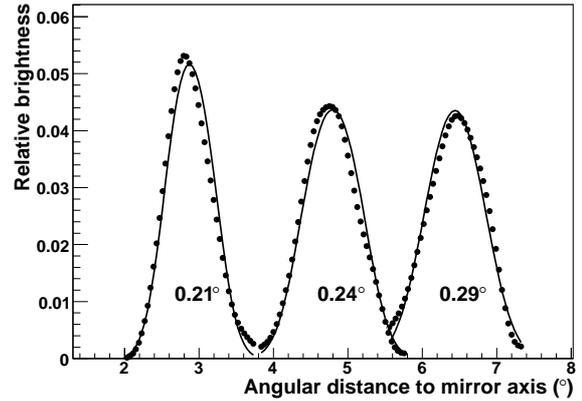}
  \caption{A star moves away from the center of the camera, and its light profile recorded by PMT is drawn in filled circles. The light profile described by a Gaussian function is drawn in a solid line. The sigma of Gaussian denotes that the spot size is also marked.} \label{spot-shape}
\end{figure}

The night sky background is measured similarly as a shower trigger is formed when the doors are open (Fig.\ref{Cherenkov-signal}). The fluctuation in the night sky back-ground is typically more than 2.2 photon electrons per 20 ns per tube on a clear moonless night.  This suggests a much stronger noise than electronic noises. One of the important sources of the sky background is light from the stars (Fig.\ref{sky-bk}), which shows the sky background at night, in which each pulse denotes stars passing though the FOV. A bright individual star can be traced when it passes through the telescopes.  Stars provide numerous stable point-like sources that typically sweep across a tube in about 4 minutes. Those ideal point-like sources can be used for multiple calibration purposes. For instance, well-known bright stars can be used as light houses to establish the pointing direction of the telescope itself, with accuracy proved to be better than $0.05^{\circ}$\cite{llma}. A bright star can be seen as a very stable point-like source at infinity, which is a perfect tool for measuring the spot size produced by the optical system. The background of a tube signal steadily increases when the light spot moves into the FOV of the tube, and steadily decreases when it moves away from the FOV, forming a light profile. The light profile can be fitted using a Gaussian function, whose sigma denotes the spot size. The spot size grows larger when the star moves away from the center of camera (Fig.\ref{spot-shape}).

\section{Summary}

The telescopes were successfully run at YBJ from August 2008 up to July 2009. Millions of coincidence events with the ARGO-YBJ experiment have been collected. The performance of the telescopes was studied using these events. The trigger rate is about 0.5 Hz in stereo mode. Moreover, the mode energy of the telescope is 40 TeV when a pure proton composition is assumed.

The features of the two WFCTA prototype telescopes are summarized as follows:

\begin{itemize}
\item a 4.7~$m^2$ spherical mirror,
\item a 16$\times$16 PMT array covers an FOV of $14^{\circ}\times16^{\circ}$ with $1^{\circ}$ pixels,
\item dual gain system for a dynamic range to 3.5 orders of magnitude,
\item DC coupling and modulized design for electronics,
\item three-level online trigger logic: single channel trigger based on S/N ratio, telescope trigger based
on pattern recognition, and event trigger for stereoscopic observation,
 \item maximized mobility and the telescope can be uplifted from $0^{\circ}$ to $60^{\circ}$ in elevation.
\end{itemize}

The absolute gains of the telescopes are calibrated using calibrated LEDs mounted at the centers of the mirrors. The systematic uncertainty of the calibration constant is about 7\%. The pixel gains are monitored on a daily basis.

\section{Acknowledgements}
This work is supported by the Chinese Academy of Sciences (0529110S13) and the Key Laboratory of Particle Astrophysics, Institute of High Energy Physics, CAS. The Knowledge Innovation Fund (H85451D0U2) of IHEP, China and the project Y0113G005C of NSFC also provide support to this study.

We are very grateful to the ARGO-YBJ Collaboration for authorizing us to use the data of the ARGO-YBJ experiment.

We also acknowledge the essential support of H. M. Zhang, W.Y. Chen, G. Yang, X.F. Yuan, C.Y. Zhao in the installation, debugging, and maintenance of the detector.




\end{document}